\documentclass{PoS}
\usepackage{textcomp,amsmath,amssymb}

\newcommand{\bea}{\begin{eqnarray}}
\newcommand{\eea}{\end{eqnarray}}
\newcommand{\ra}{\rightarrow}
\newcommand{\cL}{{\cal L}}
\newcommand{\cO}{{\cal O}}

\title{A new approach to  Naturalness in SUSY models.}

\ShortTitle{A new approach to  Naturalness.}

\author{{D. M. Ghilencea}\\ %\thanks{}\\

        CERN Theory Division, CH-1211 Geneva 23, Switzerland and\\
       Theoretical Physics Department, National Institute of Physics and Nuclear Engineering 
IFIN-HH Bucharest MG-6, Romania.\\
 E-mail: \email{dumitru.ghilencea@cern.ch}}

\abstract{We review recent results that provide a new approach to the old problem of 
naturalness  in supersymmetric models,  without  relying on  subjective definitions 
for the  fine-tuning associated  with {\it fixing}  the EW scale (to its measured value) 
in the presence of quantum corrections.  
The approach  can address in a model-independent way many questions related to this problem.
The results show that naturalness and its measure (fine-tuning) are 
an intrinsic part of the likelihood to fit the data that {\it includes} the EW scale.
One important consequence is that the additional {\it constraint} of fixing 
the  EW scale, usually not  imposed in the data fits of the models,
impacts on their overall likelihood to fit the data 
(or  $\chi^2/n_{df}$, $n_{df}$: number  of degrees of freedom). This result has
negative implications for the  viability of currently popular supersymmetric extensions 
of the Standard Model.}

\FullConference{Contribution to the proceedings of
the Corfu Summer Institute 2012,\\
"School and  Workshops on Elementary Particle Physics and Gravity",\\
		September 8-27, 2012, Corfu, Greece.\\
\noindent{\small\rm CERN-PH-TH-2013-065}}

\begin{document}

\section{Naturalness: many questions, few answers.}

The purpose of this talk is to provide a new, different perspective
to the old problem of naturalness in particle theory, based on recent results 
\cite{Ghilencea:2013hpa,gz} that we review here.  
As  pointed out long ago \cite{Susskind:1978ms},
in order to {\it fix} the electroweak (EW) 
scale in the Standard Model (SM) to its current value,
  in the presence of quantum corrections,
 a tremendous amount of ``fine'' tuning of parameters is required
(roughly 1 part in $10^{33}$), and this is thought to be {\it unnatural}.  
This tuning is just another face of the large  hierarchy  between the EW and the  Planck scales 
and indicates  when a model is unnatural,  {\it assuming} its validity  up to the Planck scale. 
However,  ``new physics'' can enter at some lower scale.
For example, additional symmetries can naturally explain such hierarchies 
\cite{thooft}. One such possibility is to build models based on 
scale (conformal) symmetry \cite{cs}, not presented here. Another possibility 
is to use  low-energy   supersymmetry, which is the case discussed below.
With supersymmetry  broken at a scale $m_{susy}$  above few TeV, 
there is however a ``remnant'' of the fine-tuning mentioned in the SM that 
worsens as $m_{susy}$ is increased, due to negative SUSY searches.
So the level of fine tuning  became a ``measure'' of the success of SUSY as 
a  {\it natural} solution to the hierarchy  problem i.e. a measure of the
naturalness of a model.

The problem is that we do not have a widely accepted
 definition  for fine tuning $\Delta$ and criteria
to decide when a  model is natural. This brings many open  questions.
The first definition  was~\cite{Ellis:1986yg}
\medskip
\bea\label{g0}
\Delta=\max_{\gamma_i}\bigg\vert
\frac{\partial \ln v}{\partial \ln \gamma_i}
\bigg\vert
,\qquad
\gamma=\{m_0, m_{1/2}, A_0, B_0,\mu_0\cdots\}.
\eea

\medskip\noindent
where $\gamma_i$ are components of the set $\gamma$ of
 SUSY-related parameters that define  the model and $v$ is the EW  scale
(higgs vev). 
This definition is  motivated  from the physics point of view, 
and measures the stability of the EW scale at the quantum level under small 
variations of $\gamma_i$, 
after SUSY is broken.
It was used in many tests for the naturalness of the models.
But an immediate question arises:
$\Delta$  depends on parameters $\gamma_i$, so
 should these parameters include  only those that SUSY introduces or also some 
non-SUSY ones as well, like nuisance variables (e.g. Yukawa couplings)? For example 
top Yukawa coupling could bring the dominant contribution to
  $\Delta$, so should we include $y_t$ on
the list of parameters in (\ref{g0})?
Further, one can also use  other definitions for $\Delta$, see \cite{castano}, or
more recently \cite{BBDDHH}, or for example:
\medskip
\bea\label{g1}
\Delta=\Big\{\sum_{\gamma_i}
\Big(\frac{\partial \ln v}{\partial \ln \gamma_i}\Big)^2\Big\}^{1/2},\qquad
\gamma=\{m_0, m_{1/2}, A_0, B_0,\mu_0\cdots\}.
\eea

\medskip\noindent
Do all these different definitions for $\Delta$ lead to similar conclusions for the
viable regions of the parameter space of the model? 
what is the right definition of fine tuning?
Another drawback is that $\Delta$ provides a local measure 
in the $\gamma_i$ space
of the quantum cancellations that fix the EW scale, while 
to compare models, one could say that a more global measure 
is desirable. How do we compare models of different $\Delta$,
over the entire parameter space?

Even if all theorists agree  on a particular
definition for $\Delta$,  then what  $\Delta$  
is acceptable for a  model to  be considered {\it natural}?
 In principle we want $\Delta$ be small, so the  EW scale is stable
when varying $\gamma_i$, in the presence of the radiative corrections; but
is  $\Delta=100$ (i.e. tuning of 1 part in 100) acceptable?  or $\Delta=10000$?  
 If no sign of TeV-scale supersymmetry is found at the LHC,
is there a value of $\Delta$ for which we should give up a
 model based on low-energy  supersymmetry?

Further  questions arise when comparing two models, 
as discussed in the following. 
Firstly, for a supersymmetric model that can fit 
the experimental data represented by a set of observables
 $O_i^{\rm ex}$ of theoretical values $O_i^{\rm th}$, one defines
\medskip
\bea
\chi^2=\sum \frac{(O_i^{\rm th} - O_i^{\rm ex})^2}{\sigma_i}
\label{chisq}
\eea

\medskip\noindent
A good $\chi^2$ fit requires that  ($n_{df}$ denotes the number of degrees of freedom)
\medskip
\bea\label{c1}
\chi^2/n_{df}\approx 1,\qquad\quad{\rm where}\qquad \quad  n_{df}\equiv n_O-n_p.
\eea

\medskip\noindent
with $n_O$ the number of observables fitted and $n_p$ the number of parameters
of the model ($\gamma_i$, etc).

 Suppose now that we have two models A and B: model A has a very good fit $\chi^2/n_{df} \approx 1$
 but fine-tuning  $\Delta$ (according to some definition) of order $\cO(1000)$. So the model
 is  likely but rather ...unnatural!   Consider that 
 model B has a good $\chi^2/n_{df}$ but slightly worse compared to model A, 
but its fine tuning much improved $\cO(10)$ (according to the same definition).
How do we decide which model of the two is better? This situation is confusing,
more so  since current data fits indeed report 
separately the value of $\chi^2/n_{df}$ and the value of $\Delta$. 
Also the possibility of having a realistic model
(i.e. good likelihood) that is unnatural, or a natural one but unrealistic
reflects a puzzling situation.
We would like to answer these questions, but since we cannot even agree
on a definition for $\Delta$, how to address them?

It is then highly desirable not to rely on  any {\it definition} for fine-tuning
and to  use more fundamental arguments (and eventually
 {\it derive} a measure for naturalness from these).
The first step to understand what happens is this:
one tunes the parameters $\gamma_i$ to fit the observables.
 This brings a $\chi^2/n_{df}$ ``cost''. 
One must also tune the  same parameters $\gamma_i$ to {\it fix} the  EW scale, 
which is central to the issue of fine tuning $\Delta$.  Clearly, there is no technical 
difference between the two tunings. That means that total $\chi^2$ and $\Delta$ 
must be technically related. But what is their relation?

Let us  think from a different perspective and provide the central idea.
While we cannot agree on a definition for fine-tuning (naturalness) in a model, we 
do know if a model is realistic or not: we 
test it against  experimental data. We do so by computing the likelihood to fit
 the data (frequentist approach) or the Bayesian probability of the model.   
These are well defined mathematical (probabilistic) tools used to test a model.
Next, recall the original goal of SUSY, central to the issue of
naturalness:  {\it fixing} the EW scale ($m_Z$) to its measured value, 
in the presence of quantum corrections. If we regard this 
as a {\it constraint} and impose it on the current likelihood to fit the data, 
the associated physical problem of  naturalness (fine-tuning) should be  
captured  by the mathematics that describes this constraint!
So one should compute the likelihood to fit the data
with this constraint imposed! Surprisingly, 
this is something not yet  done by the current methods
that compute the likelihood (or $\chi^2$)
 to fit the data in SUSY models. This is the central idea that we follow below.

Without assuming any  definition for fine-tuning or  reference to this concept,
we  show how the constraint  of fixing the EW scale  automatically 
leads to a correction factor that impacts on (worsens)  the  current 
likelihood to fit the  data. 
Once this mathematical  relation is established  one can
answer, in a model-independent way and on probabilistic grounds, 
the questions listed above.

\section{A new approach: how ``constrained'' likelihood accounts for naturalness. }

Let us then compute the likelihood to fit the data in a SUSY 
model under the mathematical
constraint of fixing the electroweak scale to its measured value,
 in the presence of quantum corrections. 
The constraint encodes the physics behind the naturalness 
problem,  and leads to a new perspective on this issue.
Firstly note that in accurate data fits
one usually performs a  fit with the following  observables 
(in a standard notation):
\bea\label{EWC}
\sin^2\theta_{eff}^{lep},\,\, 
\Gamma_Z,\,\,
\delta a_\mu,\,\,
m_h,\,m_W,\, 
\Omega_{DM},\,
BR(B\!\ra\! X_s\gamma),\,\, 
BR(B_s\!\ra\! \mu^+\!\mu^-),\,\, 
BR(B_u\!\ra\! \tau \nu),\,\,
\Delta M_{B_s}.
\eea

\medskip\noindent
However, fixing the EW scale to its accurately measured value ($m_Z^0$),
is not on this list, so the current likelihood tests of supersymmetric 
models do not  account for this effect (constraint).

For each observable $O_i$  denote the corresponding probability 
$P(O_i\vert \gamma, y)$, which  is usually taken a Gaussian. $P(O_i\vert \gamma, y)$  
depends on  parameters  $\gamma=\{\gamma_1,\gamma_2,....\}$ introduced
by SUSY  and on other parameters too, like Yukawa couplings  $y=\{y_t, y_b,.....\}$
that denote nuisance variables.  These are variables that can be removed
from the total likelihood by either maximizing it wrt to them (for $\gamma_i$
fixed) or by integrating over them wrt some measure.
Since observables $O_i$ are independent,  one multiplies 
their probability distributions to obtain a total  distribution 
$\cL({\rm data}\vert \gamma,y)$ where ``data'' stands for all observables.
 With  ``data'' fixed, $\cL$ is  a function of the sets 
$\gamma$, $y$ only, so we denote it 
as $\cL({\rm data}\vert \gamma,y)\equiv L(\gamma, y)$, with
\medskip
\bea\label{def1}
L(\gamma, y)
\!=\!\prod_j P(O_j\vert \gamma, y);
\qquad
 \gamma\!=\!%\frac{1}{v_0}
\{m_0,m_{1/2},\mu_0, m_0, A_0, B_0,\cdots\};
\qquad
y\!=\!\{y_t, y_b, \cdots\}
\eea
One then computes the corresponding $\chi^2\equiv -2\ln L$, by seeking a region of the 
parameter space where $\chi^2/n_{df}\approx 1$ at the minimum, by tuning $\gamma_j$ and $y_k$
to fit observables $O_i$. If one remains true to the original goal of SUSY,
one should impose on $L$ the additional constraint of fixing the EW scale 
to the measured value $m_Z^0$  and analyze its impact on $L$. This is what we shall do 
below. For technical details see the original papers \cite{Ghilencea:2013hpa}.

We start with the  scalar potential of supersymmetric models ($H_{1,2}$ higgs doublets):\medskip
\bea
V\!\!&=&\!\!  m_1^2 \vert H_1\vert^2
+ \! m_2^2 \vert H_2\vert^2-\! 
(m_3^2 H_1. H_2+ h.c.)+(\lambda_1/2) \vert H_1\vert^4
\!+(\lambda_2/2) \vert H_2\vert^4
\!+\lambda_3  \vert H_1\vert^2 \vert H_2\vert^2 
\nonumber\\[2pt]
& +&\!\lambda_4 \vert H_1. H_2 \vert^2
 +\big[(\lambda_5/2) (H_1. H_2)^2
 +\lambda_6 \vert H_1\vert^2 (H_1. H_2)+
\lambda_7 \vert H_2 \vert^2 (H_1. H_2)+h.c.\big]
\eea

\medskip\noindent
Some $\lambda_j$ are non-zero only at quantum level (considered
here). Introduce effective couplings $\lambda$, $m$
\medskip
\bea
\lambda& \equiv &\lambda_1/2\,\cos^4 \beta
+\lambda_2/2 \,\sin^4\beta+(\lambda_3+\lambda_4+\lambda_5)/4\,\sin^2 2\beta
+(\lambda_6\,\cos^2\beta+\lambda_7\,\sin^2\beta);
\nonumber\\[2pt]
m^2 &\equiv  & m_1^2\cos^2\beta\!+m_2^2\sin^2\beta\!-\! m_3^2\sin 2\beta,
\eea
We also denote $v^2\equiv\langle h_1^0\rangle^2+\langle h_2^0\rangle^2$.
Then  the EW minimum conditions determine  $v$  and $\tan\beta$:
\medskip
\bea
v- (-m^2/\lambda)^{1/2}=0,
\qquad {\rm and}\qquad
\tan\beta-\tan\beta_0(\gamma, y, v)=0.
\label{min2}
\eea

\medskip\noindent
where $\beta_0$ is the root of the second minimum condition. Let us denote 
\bea\label{min3}
\tilde v(\gamma, y, v, \beta)\equiv (-m^2/\lambda)^{1/2}
\eea
which at the  quantum level depends on the arguments shown.

The ``constrained'' likelihood, hereafter denoted $L_w$, to fit the data 
and also to {\it fix} the  electroweak scale to its value ($m_Z^0$) can be
written in terms of the usual likelihood   $L$  to fit 
observables {\it other than} the EW scale itself, as follows:
\medskip
\bea
 L_w(\gamma,y)\!\! &=&
m_Z^0\! \int  d v\,  d(\tan\beta)\,\, 
\delta \big[v- (-m^2/\lambda)^{1/2}\big]\,\,
\delta\big[\tan\beta-\tan\beta_0(\gamma, y, v)\big]\,\,
 \delta(m_Z-m_Z^0)\,\,
L(\gamma, y, v,\beta)\,\,
\nonumber\\[4pt]
&=&
v_0\,\,
\delta\big[v_0-\tilde v(\gamma, y,v_0,\beta_0(\gamma,y,v_0)\big]
\,\,
L\big(\gamma, y, v_0, \beta_0(\gamma,y,v_0)\big)
\nonumber\\[6pt]
&=&\frac{v_0}{\vert\nabla \tilde v\vert_o}\,\,
 \delta\big[n_i (\ln z_i-\ln z^0_i)\big]\,\,
 L\big( \gamma, y,\,v_0,\beta_0(\gamma,y,v_0)\big),
% }{\Delta_q(\gamma^0\!,y^0)},\,
\qquad {\rm with}\qquad 
%%%\tilde v\equiv -\frac{m^2}{\lambda},\,\,\,\,\,\,
z_i\!\equiv\!\{\gamma_j,y_k\}.\quad\,\,\,
 \label{oo}
 \eea

\medskip\noindent
with a sum over repeated index $i$; $m_Z^0$ in front of the integral
is a normalization factor and
\bea
m_Z=(g_1^2+g_2^2)^{1/2}\,v/2,\qquad
m_Z^0=(g_1^2+g_2^2)^{1/2}\,v_0/2,\qquad
v_0=246\, {\rm GeV}\qquad
\eea
 $m_Z$ is the theoretical mass of Z boson, $m_Z^0$ is its measured value
($\approx 91.2$ GeV)  and $g_1$, $g_2$ are gauge couplings of U(1), SU(2).
Under integral (\ref{oo}) two  Dirac delta functions 
impose EW minimum conditions (\ref{min2}).
In the presence of these functions, the integrals over  $v$, $\tan\beta$ 
are just a formal way to say that we solved the constraints of the EW minimum.
Actually, the integral over $v$ can  be regarded as an integral over $m_Z\propto v$ 
which is an observable of probability distribution
 $\delta (m_Z-m_Z^0)$; the role of this delta function is 
to fix the EW scale to the measured value. 
Using $\delta(m_Z-m_Z^0)$ is justified since $m_Z$ is accurately measured. 
In the last step in (\ref{oo}) a Taylor expansion of $\tilde v$ near $v_0$ was done,
and $n_i$ are components of the normal to the surface: 
$v_0-\tilde v\big(\gamma, y,v_0,\beta_0(\gamma,y,v_0)\big)=0$
which has  the solution
 $\gamma_j=\gamma_j^0$, $y_k=y_k^0$. So
$n_i=(\partial_i \tilde v/\vert\nabla \tilde v\vert)_o$ where subscript 
``o'' indicates evaluation  at  $\gamma_j^0$, $y_k^0$.
$\vert \nabla (...)\vert_o$  denotes the
gradient  (in parameter space $\ln\gamma_j$, $\ln y_k$),
 evaluated again at $\gamma_j^0$, $y_k^0$
  \cite{Ghilencea:2013hpa}.

From eq.(\ref{oo}), we conclude that
 the ``constrained'' likelihood $L_w$ is nonzero if 
$\gamma_j=\gamma_j^0$, $y_k=y_k^0$. These parameters are correlated by
$v_0-\tilde v(\gamma^0, y^0,v_0, \beta_0(\gamma^0,y^0,v_0))=0$  so one
could eliminate one of them (usually chosen to be $\mu_0$).
  With this in mind, the result of the last equation becomes:
\medskip
\bea\label{ttt}
L_w(\gamma^0\!, y^0)=
\frac{L\big(\gamma^0; y^0, v_0, \beta_0(\gamma^0,y^0,v_0)\big)}
{\Delta_q(\gamma^0\!\!, y^0)}.
\eea
where $\Delta_q\equiv \vert\nabla \ln\tilde v\vert_o$,  giving
\bea
\label{deltaq}
\Delta_q(\gamma^0\!\!, y^0)=\bigg\{
\sum_{z_i=\{\gamma_j, y_k\}}\!\!\!
 \Big(\frac{\partial \ln \tilde v}{\partial \ln z_i} \Big)^2_{\! o}\bigg\}^{\frac{1}{2}}
\eea

\medskip\noindent
where notice that the sum over $z_i$ extends over both $\gamma_1,\gamma_2,....$ and
 $y_t, y_b,....$ parameters\footnote{In the last step in eq.(\ref{oo}) we used $\ln z_i$ as 
fundamental variables instead of $z_i$, where $z_i=\gamma_1,\gamma_2,...,\gamma_n; y_t, y_b,...$;
this  ensures dimensionless arguments under  the last Dirac delta function. 
This equation can however
 be written in terms of $\gamma_j$ and $y_k$  after prior normalization of 
$\gamma_j$ to some scale (like $v_0$, etc), to ensure they are dimensionless.
In this case the expression of $\Delta_q$ in eq. (\ref{deltaq}) is  changed  by
replacing $\ln\gamma_i \ra \gamma_i/v_0$ and $\ln y_k \ra y_k$. }.

The result in (\ref{ttt}) shows that to maximize $L_w$ one should maximize not the traditional
likelihood $L$ (as done in present data fits),  but its  {\it ratio} to the 
quantity $\Delta_q$ that emerged above. Points in the parameter space with large 
$\Delta_q$  can reduce  the overall $L_w$.
The presence of  $\Delta_q$ in the constrained $L_w$ 
is entirely due to imposing on the traditional likelihood $L$, 
of the constraint of fixing the EW scale in the presence of quantum 
corrections. Since this constraint encodes the physics of the naturalness problem,
the quantity   $\Delta_q$ in (\ref{ttt})
can only be interpreted as a 
model-independent  measure of naturalness i.e. fine-tuning. It is for this reason
that in the following we often refer to $\Delta_q$ as ``fine-tuning''.
We  stress that $\Delta_q$  is a quantity that was {\it derived} 
when computing $L_w$,  and not an ad-hoc definition! 
$\Delta_q$ of (\ref{ttt}) depends 
 on SUSY parameters $\gamma_j$ and  on
nuisance variables like Yukawa couplings $y_k$. 
This answers the question on the parameters with respect to  which one should  actually
compute  fine-tuning.

Further, using common  assumptions such as Gaussian distributions for the observables,
then $\chi^2=-\ln L$, in which case the relation in (\ref{ttt}) can 
be rewritten as
\medskip
\bea\label{chisq2}
\chi^2_{w}=\chi^2+2\ln\, \Delta_q.
\eea

\medskip\noindent
A  SUSY model that fits well the data {\it and} also fixes 
the EW scale must have  total $\chi^2_w/n_{df}\approx 1$. 
 Until now one would only 
try to satisfy the weaker condition $\chi^2/n_{df}\approx 1$ associated with
fixing observables other than $m_Z$, and which  is not 
easily satisfied in many SUSY models.

Relation (\ref{chisq2}) for $\chi^2_w$ makes no distinction between tuning parameters to fit
 the EW scale or some other observable. Indeed,
fixing the EW scale  brings  a $\chi^2$ ``cost'' ($2\ln\Delta_q$)  for the model when  
$\Delta_q>1$, just like any other observable.
 Eq.(\ref{chisq2}) also shows that  we should actually minimize $\chi^2_w$, i.e. 
{\it the sum} of  two  contributions. A small 
$\Delta_q$ is preferable, in agreement with the traditional 
physical intuition (although it does  not have to be minimized on its own).
Finally, eq.(\ref{chisq2})  answers the old question of how to compare two models: one with 
``good'' $\chi^2/n_{df}$ but  ``bad'' fine-tuning and another with 
``not as good''  $\chi^2/n_{df}$ but ``not as bad'' fine-tuning:
one should simply minimize and compare  their {\it total}  $\chi^2_w/n_{df}$
which encodes the information of their 
likelihood to fit the data  that includes the EW scale.
In brief, naturalness is an intrinsic part of the likelihood to fit the data.

We can also derive a criterion for what is acceptable fine tuning in a SUSY
 model:  we demand a good fit for the constrained likelihood, i.e. 
$\chi^2_w/n_{df}\approx 1$ or so. This requires that
\medskip
\bea\label{eq2}
\Delta_q < \exp(n_{df}/2)
\eea

\medskip\noindent
which should be comfortably respected (given the extra $\chi^2$-''cost'' of other observables).
We thus have an answer to the question regarding an  upper bound for fine tuning,
beyond which the model may not be regarded as ``natural'' anymore. This bound 
 is general and independent of the model details other than its  number
of degrees of freedom $n_{df}$.  In many popular SUSY
 models $n_{df}$ is of order 10, meaning that an
acceptable upper bound is somewhere in the region $\Delta_q\ll \exp(5)\approx 150$.
What happens if $\Delta_q\geq \exp(n_{df}/2)$? In that case $\chi^2_w/n_{df}$ becomes 
larger than unity even before fixing observables other than the EW scale itself!
Obviously, for a given value of  $\chi^2_w/n_{df}$ such as 2 or larger we have a poor fit
 and one is unable to fit the data and (fix) the EW scale at the same time. 
Then  the model can be ruled out probabilistically.

The relation of  $\chi^2_w$ to $\Delta_q$
provided  a probabilistic interpretation
for\footnote{A probabilistic interpretation
for the inverse of $\Delta$ was suggested, on physical grounds,
 in \cite{strumia}.} $\Delta_q$.
This helps us appreciate better the significance of a given value of $\Delta_q$.
Suppose we have  $\Delta_q=10^6$ so  $\chi_w^2$ receives a correction from 
fixing the EW scale equal to $2\ln 10^6\approx 27.6$. What does this mean?
From eq.(\ref{chisq}), 
 assuming that {\it one} particular observable is $\approx 5.25 \sigma$ deviations
from its central value  would bring a correction to total $\chi^2_w$ that
is similar to that of $\Delta_q=10^6$. 
 By this argument we  have the correspondence
$\Delta_q\approx 10^6\,\,\leftrightarrow 5.25\sigma$, and similarly
$\Delta_q\approx 1000\,\,\leftrightarrow 3.5\sigma$ and 
$\Delta_q\approx 100\,\,\leftrightarrow 3\sigma$, and so on.
To see this in a broader  context,
remember that currently the disagreement between the
 muon anomalous magnetic moment in theory versus
 experiment  is about  $3\sigma$ deviations \cite{g-2}.
According to the above correspondence this is similar
to having  $\Delta_q\approx 100$. We shall discuss shortly
the numerical values of $\Delta_q$ in popular SUSY models.

\section{How the Bayesian approach accounts for naturalness.}

Here we show how to  extend our previous results to the Bayesian approach to 
fit the data, based on  \cite{gz} (section 2) reviewed here.
In this approach, one computes the Bayesian probability (``evidence'')
  $p({\rm data})$ of  a model, given the data.
Here $p({\rm data})$ is an integral over the parameter space 
$\gamma_i$ (and eventually  $y_k$ too) of the product between the
likelihood $L$ and the initial probability  distributions $p(\gamma)$ (``priors'') for the
parameters 
\medskip
\bea\label{3.1}
p({\rm data})=\int d\gamma_1.... d\gamma_n \,\, p(\gamma)\,\,L(\gamma,y,v,\beta)
\eea

\medskip\noindent
This result for 
 $p({\rm data})$ is used to compare the relative probability (viability) of
 different models. 

To extend our results we simply use  (\ref{3.1}) with the replacement $L\rightarrow L_w$.
Eq.(\ref{oo})   that imposes the constraint of fixing the EW scale,
 is multiplied  by  $p(\gamma)$  and  integrated  over $\gamma_i$. 
The result is\footnote{One can also use that 
(the surface integral is over $S$ defined by $g=0$):
\bea
\int_{R^n}
f(z_1,...,z_n)\, \delta(g(z_1,...,z_n))\,dz_1....dz_n=
\int_{S_{n-1}} dS_{n-1}\,f(z_1,...,z_n)\,\frac{1}{\vert \nabla_{z_i} g\vert}
\eea}
\medskip
\bea\label{bayesian}
p({\rm data})&=&
m_Z^0\!\int\!
d\gamma_1....d\gamma_n\,\, p(\gamma)\,\, dv\,\, d(\tan\beta)\,
\nonumber\\[4pt]
&&\,\,\,\, \times\,\,\delta\big[v-\tilde v(\gamma, y, v, \beta)\big]
\, \,\delta\big[\tan\beta-\tan\beta_0(\gamma, y, v)\big]
\,\, \delta(m_Z\!-\!m_Z^0) \,\,\, L(\gamma, y, v,\beta)
\\[4pt]
&=&v_0\int  d\gamma_1....d\gamma_n\,\,p(\gamma,y)\,\,
\delta\big[v_0-\tilde v(\gamma, y,v_0,\beta_0(\gamma,y,v_0))\big]
\,\,\, L\big(\gamma, y, v_0, \beta_0(\gamma,y,v_0)\big)
\nonumber\\
&=&\int_{\Sigma}
 dS_{\ln\gamma}\,\,\frac{1}{\vert \nabla_{\ln\gamma}\,\ln\tilde v\vert}
\,\,\, L\big(\gamma,y,v_0,\beta_0(\gamma,y)\big)
\,\gamma_1...\gamma_n\,\, p(\gamma),
\qquad
 \Sigma:\, v_0=\tilde v\big(\gamma,y,v_0,\beta_0(\gamma,y,v_0)\big)\nonumber
\eea

\medskip\noindent
This is just a {\it global} version in parameter space of the local  result in
 (\ref{oo}), (\ref{ttt}); in deriving this result,  the priors are spectators.
 The integral above is over
the surface $\Sigma$ defined by $v_0-\tilde v(\gamma, y, v_0, \beta_0(\gamma, y, v_0))=0$. 
Note the emergence of the suppression factor
 $\vert\nabla_{\ln\gamma}\ln\tilde v\vert=\Delta_q$ where $\Delta_q$ 
is that given by eq.(\ref{deltaq}) but with the sum 
restricted to  $\gamma$ parameters ($\nabla_{\ln\gamma}$ denotes the gradient
computed wrt $\ln\gamma_j$ variables).
If one integrated (\ref{bayesian}) over Yukawa (nuisance) parameters too, then  $\Delta_q$ 
is exactly that of eq.(\ref{deltaq}) \cite{gz}.
The  $1/\Delta_q$ factor  under the integral of $p({\rm data})$ 
in the last line of (\ref{bayesian})
 acts as an extra  prior and is solely due to the constraint
 of fixing the EW scale via $\delta(m_Z-m_Z^0)$, as also seen in the local case.
By comparing this result for $p({\rm data})$, evaluated numerically for different models,
one  decides which model is more likely,
while taking into account  the effect of $1/\Delta_q$ associated with naturalness.
Finally, eq.(\ref{bayesian}) simplifies further for 
 logarithmic priors, $p(\gamma)=1/(\gamma_1\gamma_2....\gamma_n)$.

Let us mention an  equivalent form of the result in (\ref{bayesian}) which is
\medskip
\bea
p({\rm data})=
\int_{\Sigma}  dS_\gamma\,\,\frac{1}{\vert \nabla_\gamma\,\ln\tilde v\vert}
\,\,\, L\big[\gamma,y,v_0,\beta_0(\gamma,y,v_0)\big]\,p(\gamma)
\eea

\medskip\noindent
 where the gradient $\nabla_{\gamma}$ and  $d S_\gamma$ are defined 
with respect to parameters $\gamma_i$ instead of $\ln\gamma_i$ used in (\ref{bayesian}). In this
case $\Delta_q$ changes accordingly, $\Delta_q=\vert \nabla_\gamma\,\ln\tilde v\vert$.

Unlike the ``frequentist'' approach  discussed in the previous section,
current data fits in the Bayesian approach already include the ``cost'' of naturalness 
in their results.
Early attempts were based on a choice of priors that go like $1/\Delta$ where $\Delta$ was
an  ad-hoc definition for fine-tuning  wrt a given parameter \cite{ben}. 
That meant that points that were badly fine-tuned wrt that particular choice of fine-tuning,
were giving a smaller contribution to  global $p({\rm data})$.
More recently, a so-called ``Jacobian factor'' was identified to account
for a similar effect  \cite{bc,Fichet}, starting from
assumptions  similar to those used  here (see \cite{Fichet} for more details).
It would be interesting to establish 
the exact relation of this Jacobian factor to  the factor $1/\Delta_q$ that 
we identified  above to  account for naturalness. This ends our discussion on
the Bayesian approach and on how $\Delta_q$ emerges in this case.

\section{The minimal value of $\Delta_q$ 
in supersymmetric models and its impact on $\chi^2_w/n_{df}$.}

\begin{figure}[t!]
\vspace{-0.4cm}
\begin{center}
\includegraphics[width=6.7cm,height=4.5cm]{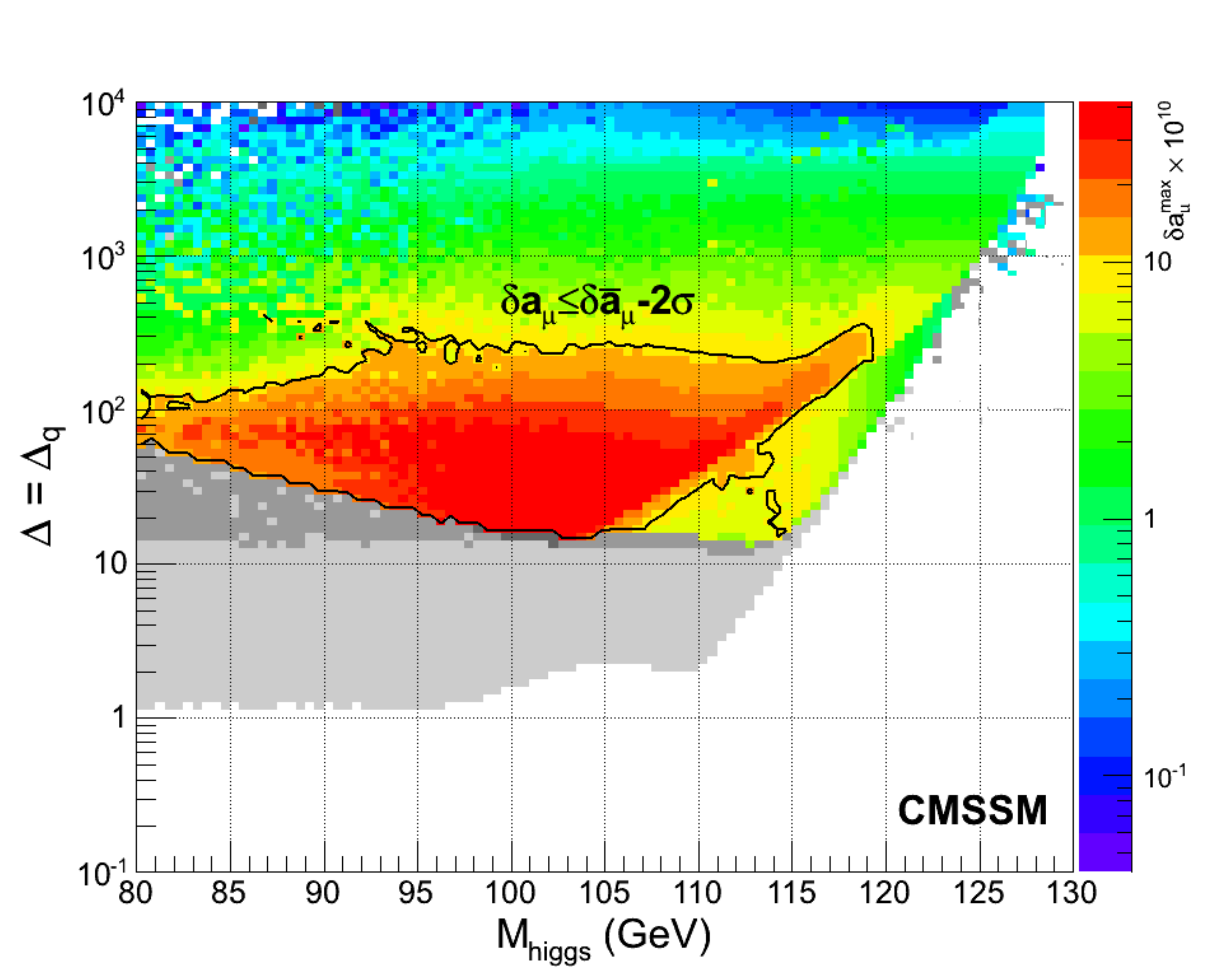}
\includegraphics[width=6.7cm,height=4.5cm]{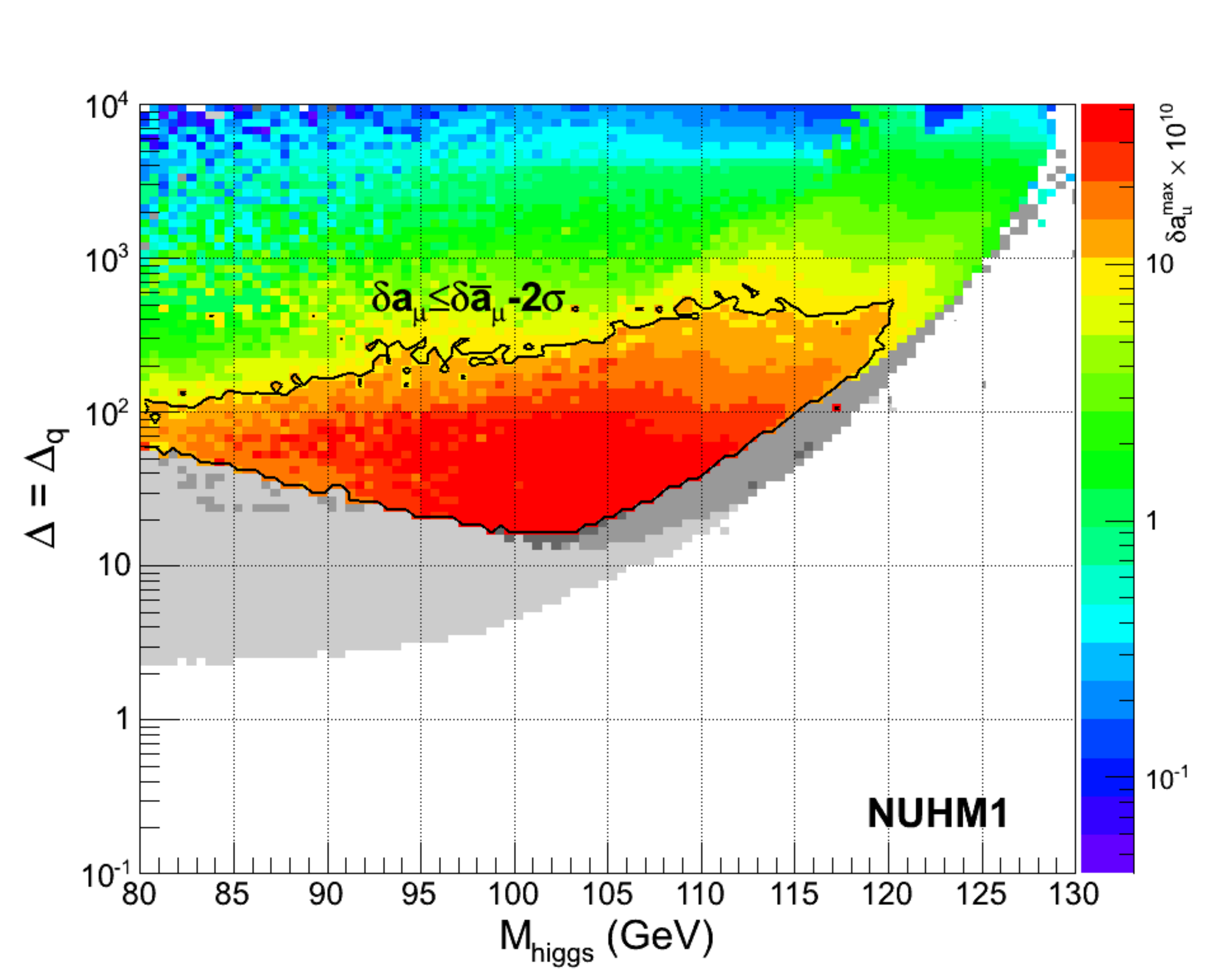}
\includegraphics[width=6.7cm,height=4.5cm]{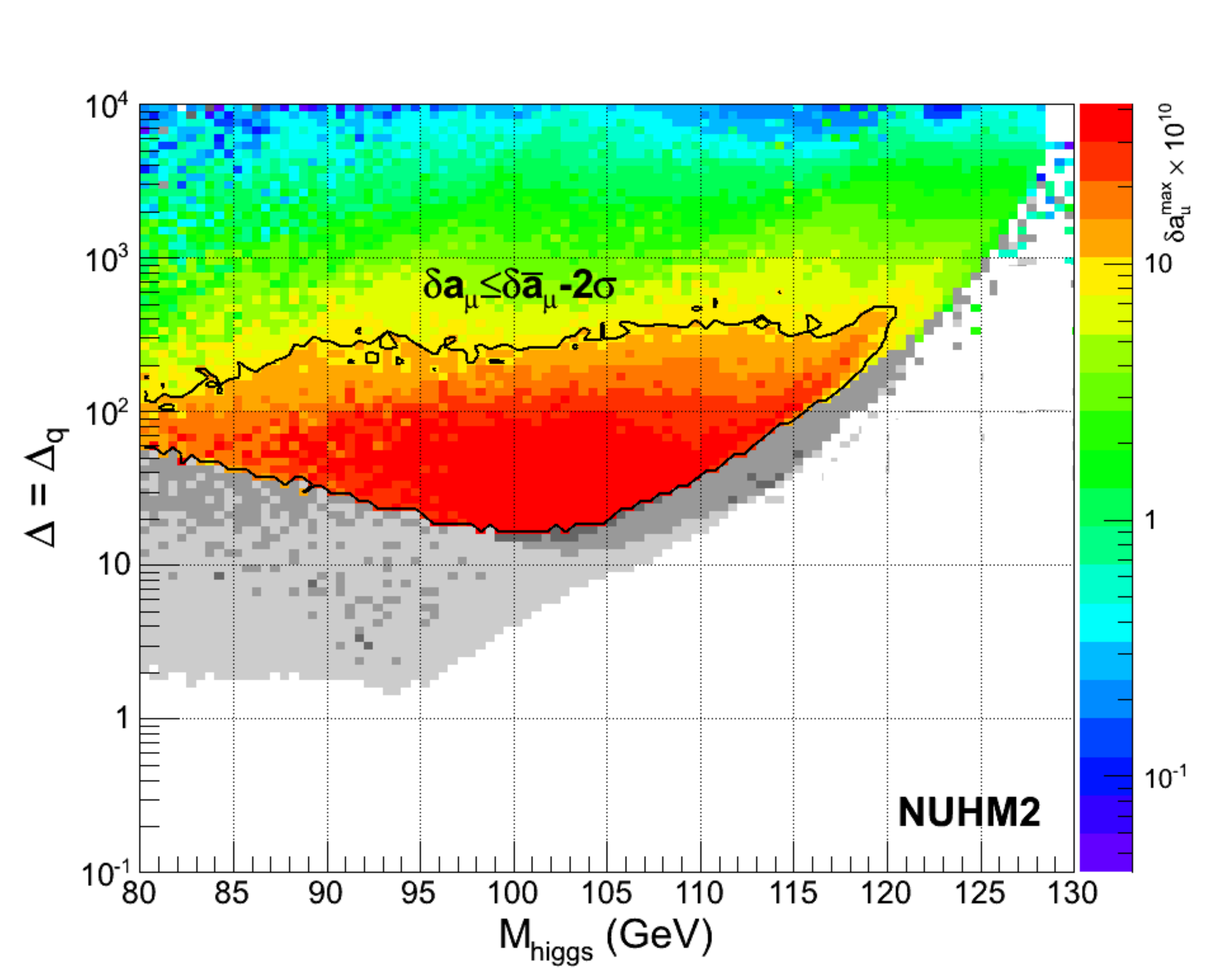}
\includegraphics[width=6.7cm,height=4.5cm]{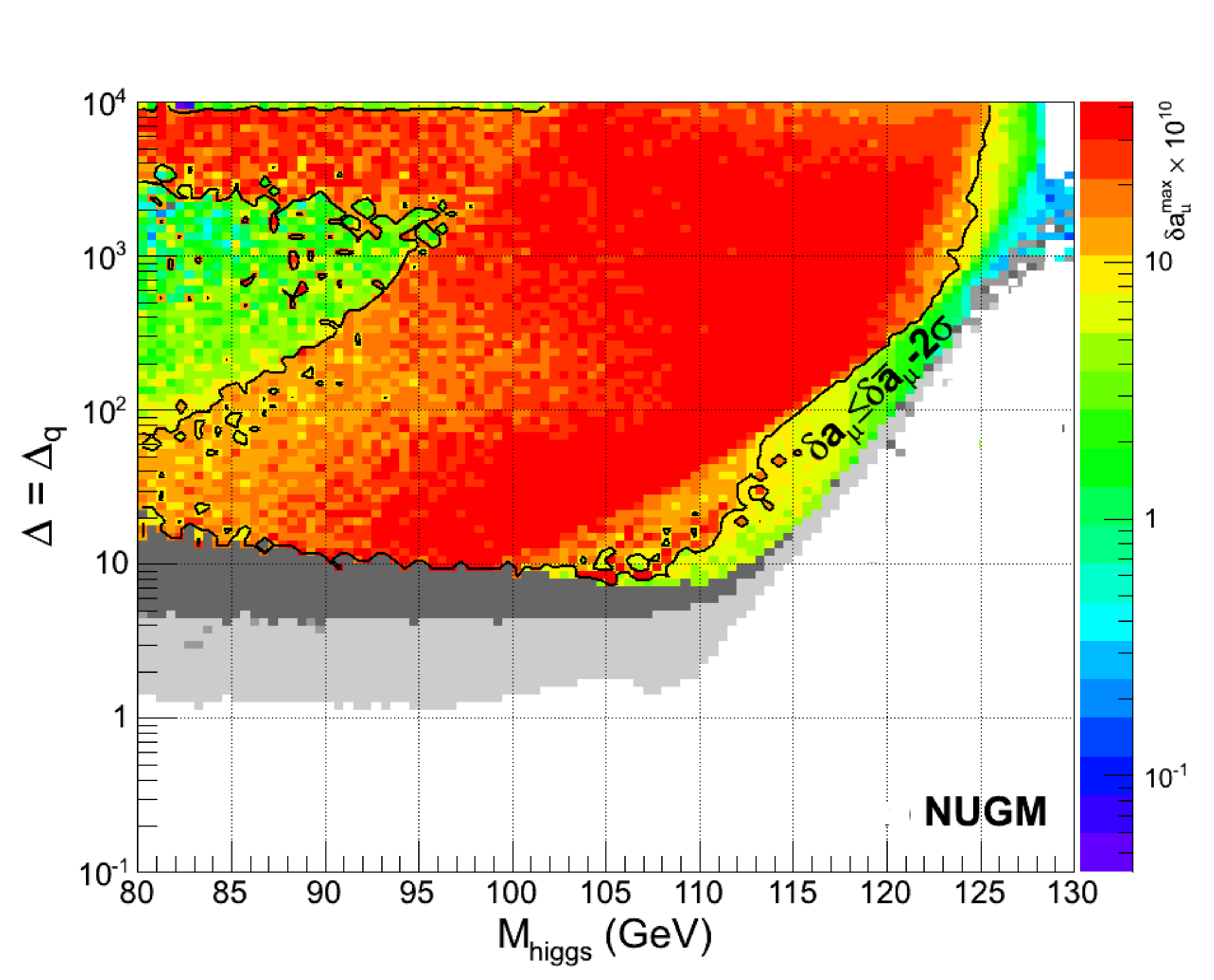}
\end{center}
\renewcommand{\baselinestretch}{0.9}
\vspace{-0.2cm}
\caption{\small 
$\Delta_q$ versus $M_{higgs}$ in various models,  from ref.\cite{gz}; 
lightest grey (0)  area: 
excluded by SUSY mass bounds; 
darker grey (1): excluded by $b\!\rightarrow \! s \gamma$, 
$B\!\rightarrow\! \mu^+\mu^-$, $\delta\rho$; dark grey (2): excluded 
by condition  $\delta a_\mu \geq 0$.
Area inside the closed contour: allowed by data and
with $2\sigma$ deviation of $g-2$:
 $\delta a_\mu\leq (25.5 + 2\!\times\! 8) 10^{-10}$;
$\delta a_\mu^{max}$ is shown colour encoded.
Area outside closed  contour: $\delta a_\mu^{max}\!\leq\! (25.5 - 2\!\times\! 8) 10^{-10}$
($2\sigma$). Only in the NUGM   is one close to satisfying the $g-2$ constraint
within $2\sigma$, not too surprising given its non-universal 
gaugino masses. In all plots the dark matter relic density
was computed \cite{micromegas} and can be fitted within $3\sigma$ \cite{gz}. For similar plots in 
NMSSM or GNMSSM see \cite{Ross:2012nr} while for the GNMSSM in the limit of a 
 massive gauge singlet (integrated out) see plots in \cite{Cassel:2009ps}.}
\label{figure1}
\end{figure}

The next step   in this analysis  is to answer what 
values $\Delta_q$ of  eq.(\ref{deltaq})
 takes in current SUSY models.
 We  then  examine the impact it has on total $\chi^2_w/n_{df}$, see eq.(\ref{chisq2}).
Can one  satisfy  condition (\ref{eq2})? 
To this purposes we use the  results in  \cite{gz}. For related
studies see 
\cite{Ross:2012nr,Cassel:2009ps,Kowalska:2012gs,Cassel:2010px,ross,finetuning,giudice}
and for recent data fits in  SUSY models  see \cite{Kowalska:2012gs,recent}.  
We  use  eq.(\ref{deltaq}) evaluated wrt to the SUSY parameters
$\gamma_i$ only, listed below for each model considered;
this restriction  underestimates  $\Delta_q$, so
from this point of view our analysis is conservative.
The models considered here are~\cite{AbdusSalam:2011fc}:

\medskip\noindent
$\bullet$ the constrained MSSM model (CMSSM), 
of parameters $\gamma_j \equiv \{m_0, m_{1/2}, \mu_0, A_0, B_0\}$,
in a standard notation for soft masses of squarks and sleptons ($m_0$),
gaugino ($m_{1/2}$), usual $\mu$ term ($\mu_0$), trilinear ($A_0$) and bilinear ($B_0$)
SUSY-breaking terms. 

\noindent
$\bullet$ the NUHM1 model:  this is a CMSSM-like model but with  Higgs
 masses in the ultraviolet (uv) different from $m_0$, 
$m_{h_1}^{uv}=m_{h_2}^{uv}\not= m_0$, with parameters
$\gamma_j\equiv \{m_0, m_{1/2}, \mu_0, A_0, B_0, m_{h_1}^{uv} \}$.

\noindent
$\bullet$ the NUHM2 model: this is a CMSSM-like model with non-universal Higgs mass, 
$m_{h_1}^{uv}\!\not=\!m_{h_2}^{uv}\!\not=\! m_0$,  
with independent parameters  $\gamma_j\equiv 
\{m_0, m_{1/2}, \mu_0, A_0, B_0, m_{h_1}^{uv},m_{h_2}^{uv}\}$.

\noindent
$\bullet$ the NUGM model:
 this is a CMSSM-like model with non-universal gaugino masses 
$m_{\lambda_i}$, $i=1,2,3$,  with 
$\gamma_j=\{m_0, \mu_0, A_0, B_0, m_{\lambda_1}$, $m_{\lambda_2}, m_{\lambda_3}\}$.

\noindent
$\bullet$ the NMSSM model: compared to the CMSSM, in this case
there is an additional gauge singlet, so
$\gamma_j=\{m_0, m_{1/2}, \mu_0, A_0, B_0, m_{1/2},m_S\}$ where $m_S$ is the singlet
soft mass.

\noindent
$\bullet$ the general NMSSM model (GNMSSM): this is an extension of the NMSSM
with a  bilinear superpotential term, $M S^2$, where $S$ is the singlet
superfield. So we have one more parameter ($M$) 
in addition to the parameters of the NMSSM.

\medskip\noindent
In all models a complete scan over the entire parameter space $\gamma_i$
is performed \cite{gz} including $\tan\beta$  with standard experimental constraints\footnote{
For the experimental data used see table~1 of \cite{gz}.
The muon $\delta a_\mu$  is not imposed but  shown separately in Figures~1.}. 
The result for $\Delta_q$ 
(computed at two-lop  using SOFTSUSY \cite{softsusy}), is plotted in Figure~1
as a function  of the SM-like  higgs mass $m_h$ (computed at two-loop leading log),
for each model, except the NMSSM, GNMSSM cases.
Notice the log scale in the plots (on OY axis),
indicating an approximately exponential increase of $\Delta_q$ wrt $m_h$
 \cite{Cassel:2010px}. 
We find it interesting that if one also computed 
$\Delta$ defined by (\ref{g0}), the plots have an identical form to those for $\Delta_q$,
but only slightly shifted  along OY axis,
to a lower $\Delta$ by a factor between 1 and 2.

\begin{table}[!t]
\centering
\begin{tabular}{lcrrlrlcc}
\hline\hline\\[-7pt]
Model  &  $n_p$  &  Approx  
& $\Delta_q${\footnotesize (115)} &  $\delta\chi^2${\footnotesize (115)} 
& $\Delta_q${\footnotesize (126)} &  $\delta\chi^2${\footnotesize (126)}
&  $n_{df}$ &  $\chi^2_w/n_{df}${\footnotesize (126)}\\[5pt]
\hline\\[-7pt]
CMSSM  &    5      &  2-loop  &    15  &  5.42  &  1800  & 14.99  &  9  & 2.66 \\[0.5ex]

NUHM1  &    6      &  2-loop  &   100  &  9.21  &  1500  & 14.63  &  8  & 2.83 \\[0.5ex]

NUHM2  &    7      &  2-loop  &    85  &  8.89  &  1300  & 14.34  &  7  & 3.05 \\[0.5ex]

NUGM   &    7      &  2-loop  &    15  &  5.42  &  1000  & 13.82  &  7  & 2.97 \\[0.5ex]

NMSSM  &    6      &  1-loop  &    12  &  4.97  & $>$200 & 10.59  &  8  & 2.32\\[0.5ex]

GNMSSM &    7      &  1-loop  &     12 &  4.97  &   27 & \,\,\,6.59  &  7 & 1.94 \\[1ex]
\hline 
\end{tabular}
\\[1ex]
\caption{\small
The correction $\delta\chi^2\equiv 2\ln\Delta_q(\gamma)$,
 the number of parameters $n_p$ and degrees of freedom $n_{df}$ in  SUSY
 models, for $m_h\approx 115$ GeV corresponding to the
 pre-LHC  bound  and for $m_h\approx 126$ GeV. 
For 115 GeV, one had $\delta\chi^2/n_{df}<1$, thus
leaving the possibility of having $\chi^2_w/n_{df}\approx 1$.
For the 126 GeV value, $\delta\chi^2/n_{df}>1$ in all models above, except the 
GNMSSM where this is close to unity. This means that the requirement of a good fit
($\chi^2_w/n_{df}\approx 1$) is already violated by the fixing of the EW scale alone,
before considering the additional $\chi^2$ cost of other observables. Assuming
these give the usual $\chi^2/n_{df}\approx 1$ i.e. a good fit, adding the impact of $\delta\chi^2$
leads to the total $\chi^2_w/n_{df}$ displayed in the last column.
 $n_{df}$ may vary, depending on the exact number of observables fitted.
The numerical values of $\Delta_q$ are from \cite{gz} (first 4 models) and
correspond to the plots in Figure~1, while for GNMSSM and NMSSM we used the results
in  \cite{Ross:2012nr} and \cite{Kowalska:2012gs}.}
\label{tabel1}
\end{table}

From Figure~1, after fixing the higgs mass to its measured value  $m_h\approx 126$ GeV
 at the LHC \cite{SMH} one immediately identifies  in each model the
{\it minimal} value of $\Delta_q$ obtained after scanning the whole parameter
space. The results are summarized in Table~1 for  minimal $\Delta_q$ and its corresponding
$\delta\chi^2=2\ln\Delta_q$ contribution, for $m_h$ near the pre-LHC bound (115 GeV) and 
for the current value (126 GeV).
The results in this table include additional numerical results for the 
NMSSM and also GNMSSM models  for which we did not display the 
 plots of $\Delta_q$   as a function of $m_h$, but 
used instead the results of  \cite{Ross:2012nr}, \cite{Kowalska:2012gs}.
 Although not shown in the table, for all models
$\delta\chi^2$ shown varies by $\approx 1$  for a similar sign
 variation of the higgs mass  by 1 GeV.

The results in Table~1 show  that $\Delta_q$ has a significant impact on the overall
quality of the fit (i.e. the value of $\chi^2_w/n_{df}$) since it brings 
corrections   $\delta\chi^2/n_{df}>1$ for models other than the GNMSSM.
 This means that the requirement of a good fit
($\chi^2_w/n_{df}\approx 1$) is already violated by the fixing of the EW scale alone,
before considering the additional $\chi^2$ cost of other observables.
So  naturalness,
encoded by the constraint  of fixing the EW scale,  has
a very significant $\chi^2_w$ cost. Assuming (ideally) a good fit under 
current experimental
constraints, i.e. $\chi^2/n_{df}\approx 1$, one finds that overall $\chi^2_w/n_{df}>2.3$
in the models discussed.
An exception is the GNMSSM model,  where this result is improved mildly 
($\approx 1.94$).

There is another interesting aspect in Table~1.
While in some models $\Delta_q$ at 126 GeV decreases relative 
to the CMSSM case, 
the correction $\delta\chi^2/n_{df}$ and total $\chi^2_w/n_{df}$ can actually 
increase (!), because of the stronger effect of  decreasing $n_{df}$
(larger $n_p$). Thus, unless it is very significant, a reduction of  
$\Delta_q$ due to new physics (more parameters) 
does not necessarily  lead to a reduction of $\chi^2_w/n_{df}$ and a better fit. 
In GNMSSM the reduction of $\Delta_q$ is significant enough and 
$\chi^2_w/n_{df}$ is improved relative to the CMSSM (for a recent
GNMSSM study see \cite{Dreiner}).

\section{Conclusions}

Our conclusions are best summarized by listing
the old, unanswered questions in the
  introduction on the problem of fine-tuning
in SUSY models, together  with the answers that we found. 

\medskip
\noindent
$\bullet$ Is there a relation between naturalness 
and the likelihood to fit the data, and if so, can we {\it derive}, 
on mathematical grounds, an expression for fine-tuning?

\noindent
To answer this, one should recall the original goal of SUSY,
 central to the physical problem of naturalness 
(fine-tuning): to {\it fix} the EW scale to its measured value,
in the presence of quantum corrections (while assuming the model is valid
up to the Planck scale). If we regard this as a {\it constraint} and impose
it on the current likelihood to fit the data 
 the associated physical problem
of naturalness (fine-tuning) should be captured by the mathematics that 
describes the constraint.
We computed the  ``constrained'' likelihood $L_w$ by imposing  this constraint
on the likelihood $L$ to fit the data (other than the EW scale) and found that 
$L_w=L/\Delta_q$. So one should  maximize $L_w$, not $L$. 
The factor $\Delta_q$  is 
due to  this constraint alone, so it encodes the mathematical effect of 
the naturalness; thus, it can only be regarded as a 
 {\it derived}  fine-tuning expression. 
Rather surprisingly, current data fits (in the frequentist approach) do
no account for this (suppressing) effect  on the likelihood, thus leading 
to more optimistic results.

\medskip
\noindent
$\bullet$ What are the parameters with respect to which $\Delta_q$ should be computed?

\noindent
In the literature it is often unclear whether  to include Yukawa couplings 
among these parameters. 
If one computes the constrained likelihood $L_w$ as explained, there is no
 choice: these parameters include both
the SUSY parameters and the nuisance variables  (Yukawa couplings, etc) as well.
One can maximize $L_w$ wrt Yukawa (SUSY parameters fixed) and thus eliminate them.
Alternatively, one can marginalize $L_w$ over  nuisance variables
 in which case $\Delta_q$ is computed with respect to the SUSY parameters only.

\medskip\noindent
$\bullet$ What is an ``acceptable'' value  for $\Delta_q$ identified above?

\noindent
The relation between the likelihoods $L_w=L/\Delta_q$, has a correspondent 
in $\chi^2_w=\chi^2+2\ln\Delta_q$ (assuming $\chi^2=-2\ln L$).
So total  $\chi^2_w$ accounts for  tuning  the parameters to fix 
both the usual observables {\it and} the EW scale, and it makes  no distinction
between these tunings which are treated on equal footing, as it should be the  case.
In particular, electroweak  scale (``fine'') tuning is nothing but $\chi^2$ ``cost''!
In other words, naturalness is an intrinsic part of the likelihood to fit the data
that includes the EW scale.
For a good fit one should then have a  {\it total} $\chi^2_w/n_{df}\approx 1$
where $n_{df}$ is the number of degrees of freedom.
This implies a necessary bound  $\Delta_q< \exp(n_{df}/2)$, which should be comfortably 
respected (i.e. $\Delta_q\ll \exp(n_{df}/2)$, since often $\chi^2/n_{df}$ is itself close to unity).
For the  popular  models discussed this means $\Delta\ll 90$ ($n_{df}\leq 9$).

\medskip\noindent
$\bullet$ Can we rule out supersymmetric models based on the size of the quantity
$\Delta_q$ alone?

\noindent
The answer is yes; when $\Delta_q\gg \exp(n_{df}/2)$ then the total
 $\chi^2_w/n_{df}$
is so poor that the  model cannot fit the data anymore and is ruled out. 
Note however that the addition of new physics,  that can actually 
be beyond the LHC reach, could  reduce $\Delta_q$, as seen for the 
``general NMSSM'' (GNMSSM) model.

\medskip\noindent
$\bullet$ Does $\Delta_q$ have a physical meaning? 

\noindent
The relation 
 $\chi^2_w=\chi^2+2\ln\Delta_q$ gives $\Delta_q$ a probabilistic interpretation. 
A value of say $\Delta_q=10^6$ has an effect  on $\chi^2_w$
 similar to one  observable that  
contributes to $\chi^2_w$ being $\approx 5.25\sigma$ deviations from the central  value.
By the same rule,  $\Delta_q=100$ is equivalent to a deviation of $\approx 3\sigma$,
which is the order of deviation of  the muon magnetic moment.

\medskip\noindent
$\bullet$ How can we compare two models and decide which is better: one that has a 
very good $\chi^2$ fit in the  traditional sense,  
but significant fine tuning  and one that has a nearly-as-good  $\chi^2$ fit but 
less fine-tuning ? 

\noindent
The question arises since
accurate data fits usually report separately $\chi^2/n_{df}$ and fine tuning $\Delta$ 
computed according to some {\it definition}.  We showed that to compare such  models,
one should actually   compare their  total $\chi^2_w/n_{df}$, and this answers the question above.
This is the `frequentist' case.  In the Bayesian approach  one compares
instead the ``constrained'' probabilities of the two models $p({\rm data})$
that contain an  additional  $1/\Delta_q$ factor under their integral.
This factor again  emerges   (as an extra prior) from 
the naturalness  constraint of fixing the EW scale at quantum level.

\medskip\noindent
$\bullet$ How can one reduce $\Delta_q$  and  also  $\chi^2_w/n_{df}$ to improve the data
fits of SUSY models?

\noindent
The easiest way to reduce $\Delta_q$ is to increase the effective quartic higgs coupling $\lambda$,
whose  small value is at the origin of large $\Delta_q$ in most SUSY models\footnote{For details on
this issue, see first reference in \cite{Cassel:2010px}.}.
One can increase $\lambda$
by new physics (supersymmetric!) beyond the MSSM higgs sector \cite{carena}
and possibly beyond the LHC reach too! The new physics  can be represented by
 additional massive states that couple to the higgs sector: 
 singlets, SU(2) doublets, additional massive $U(1)$ boson, etc.
However, adding new physics increases the number of parameters (reduces $n_{df}$).
Therefore, unless  the reduction in $\Delta_q$ is very significant (to 
$\Delta_q<10$ or so), the   total  $\chi^2_w/n_{df}$ can actually  increase 
due to the  reduced $n_{df}$  and to the mild (log) dependence of $\chi^2_w$ on $\Delta_q$.
Of all models discussed (MSSM or NMSSM-like) only the  GNMSSM model had $\Delta_q<10$;
this model emerges as the one with the 
best chance of a good fit ($\chi^2_w/n_{df}$) in the sense discussed above.

\medskip\noindent
$\bullet$ Can one compare the relative viability of two 
models based on the values of  $\Delta_q$ alone? 

\noindent
The answer is  negative.  
It  is the total $\chi^2_w/n_{df}$ that matters when comparing models, not $\Delta_q$ alone.
This is because  one model can have a smaller  $\Delta_q$ (by adding more parameters) 
 but a larger   $\chi^2_w/n_{df}$  than  in the second model (see previous point).

\medskip
To conclude,  naturalness in SUSY models is an intrinsic part of the likelihood to fit 
the experimental  data that includes the EW scale. This helps address many open questions 
on this issue, as shown above. The results above  indicate 
that for some popular SUSY  models  the value obtained for the corrected  likelihood 
(i.e.  $\chi^2_w/n_{df}$) already questions their  viability, except perhaps the case of 
GNMSSM.  This can be  
investigated further by  performing  accurate data fits that evaluate the  new
$\chi^2_w/n_{df}$ that we identified.

\bigskip\bigskip\noindent
{\bf Acknowledgements:}
 The author thanks Hyun Min Lee, Myeonghun  Park and Graham Ross,
for their  collaboration on the original papers reviewed here (ref. \cite{Ghilencea:2013hpa,gz}).
This work was supported by a grant of the Romanian National Authority for Scientific
Research, CNCS - UEFISCDI, project number PN-II-ID-PCE-2011-3-0607.

\end{document}